\documentclass[a4,12pt]{article}
\usepackage{subcaption}
\usepackage{multirow}
\usepackage{adjustbox}
\usepackage{caption}
\usepackage{graphicx}
\usepackage{amsmath}
\usepackage{amsfonts}
\usepackage{amssymb}
\usepackage{xcolor}
\usepackage{pdfcomment}
\usepackage{hyperref}
\usepackage[sort&compress,square,numbers,comma]{natbib}
\usepackage{mathtools}
\usepackage{float}
\usepackage{authblk}
\usepackage{enumitem}
\usepackage[pass=paperwidth=8.76in,paperheight=11in]{geometry}

\titlepage

\begin{document}
\title{Stability of neutrino oscillation parameters at low
energy scale with the variations of SUSY breaking
scale under Renormalisation Group Equations}
\date{}
\author[1]{Kh. Helensana Devi \footnote{helensanakhumanthem2@manipuruniv.ac.in}}
\author[1]{K. Sashikanta Singh \footnote{ksm1skynet@gmail.com}}
\author[1,2]{N.Nimai Singh \footnote{nimai03@yahoo.com} }
\affil[1]{\small{{Department of Physics, Manipur University, Imphal-795003, India} }}
\affil[2]{Research Institute of Science and Technology, Imphal-795003, India }

\maketitle \thispagestyle{empty}
\begin{abstract}
We discuss the stability of the neutrino oscillation parameters at low energy scale including self-complementarity (SC) relations among mixing angles under radiative corrections with the variation of SUSY breaking scale ($m_s$) in both normal and inverted hierarchical cases. We observe that the neutrino oscillation parameters including the SC relation maintains  stability  at the electroweak scale within $1\sigma$ range of the latest global fit data. NH case maintains more stability than IH case. All the numerical values related to the absolute neutrino masses viz., $\Sigma |m_i|$, $m_{\beta}$ and $m_{ \beta \beta}$ are found to lie below the observational upper bound.

\end{abstract}

\titlepage

\section{Introduction} 
Neutrino physics has been greatly expanded in the
past decades, especially due to neutrino
oscillation experiments, the neutrinoless double beta decay, tritium beta decay together with cosmological observations. The Double Chooz \cite{chooz} and Daya Bay \cite{daya} experiments measured non-zero value of $\theta_{13}$ with enough precision to open the possibility of a Dirac CP phase $\delta_{CP}$ in the mixing of the leptonic sector of the SM described through the Pontecorvo-Maki-Nakagawa-Sakata (PMNS) matrix \cite{pmns,pmns4}. PMNS matrix, describes the relation that the three known neutrino flavor states are mixtures of three mass states \cite{pmns,pmns2}, which can be parameterized by three mixing angles, and Dirac CP phase $\delta_{CP}$.

From current neutrino oscillation data, we notice that the neutrino oscillations accord with positive information about non-zero neutrino masses which will be highly preceded to provide clues about the origin of neutrino masses and lepton flavor. The question of neutrino mass hierarchy, which requires experimental and theoretical effort, is a central question in particle physics today. Current experiments and future investigations under construction are aimed to determine the missing neutrino parameters, such as the CP-violating phase, the absolute neutrino mass scale, and the sign of the mass-squared difference for neutrinos.

Neutrino mass models are extensively studied in the supersymmetric framework which exists in literature only. However, Large Hadron Collider (LHC) aims to reveal the physics beyond the Standard Model with center-of-mass collision energies of up to 14 TeV to check the presence of supersymmetric particles (third
run of LHC reaches 13.6 TeV slightly higher then 13 TeV of the second run \cite{eli}). As of now there is no hint on the existence of supersymmetric particles.

 The current  neutrino oscillation data indicates that the self-complementarity (SC) among three mixing angles can be satisfied within $3\sigma$ ranges. Hence it may lead to investigate a flavor physics beyond the SM, and to check its stability is also necessary. We assume a self-complementarity (SC) relation for mixing angles, $\theta_{23} = q(\theta_{12}+\theta_{13})$ \cite{scr,scm,scs,sczx,scz,qlcsc1,sczz,scn2}, taking q = 1.11, which is phenomenologically analogous to a relation for quark and lepton mixing angles known as Quark-Lepton Complementarity (QLC) relations, $\theta_{sun} + \theta_{C} = \pi/4$ \cite{qlcsc1,scr,scm,scs} between the leptonic 1-2 mixing angle $\theta_{sun}$ and the Cabibbo angle $\theta_{C}$. We apply the renormalization group evolution of parameters to compare the predicted theoretical values at high energy scale with experimentally observed low energy values.
 
  In the present work, there are 9 free parameters (3 mixing angles, 3 phases,
and 3 neutrino masses). Since it is so complicated to deal with all 9 parameters as completely free ones, we therefore try to reduce the free parameters. We will analyze both  Normal Hierarchy (NH) and Inverted Hierarchy (IH) cases. We assume another SC relation for neutrino masses. For NH, we assume  a simple relation among neutrino masses, $m_2 = 1.236 \times m_1$ and  $m_3 = 4.395 \times m_1$, taking $m_1$ = 0.012 eV, whereas in the case of IH, we assume  $m_1 = 5.686 \times m_3$ and $m_2 = 5.731 \times m_3$, taking $m_3 =0.009$ eV.

  The present investigation is a continuation of our previous work on neutrino masses and mixings with varying SUSY breaking scale in the running of RGEs \cite{rgg1,rgg2,rgg3,rgg4,rgg5,da,dac,hel}. The present work is organized as follows. In the next section, we discuss briefly on renormalization group equations. In section 3, we study the numerical analysis  and results. Our discussion and conclusions are given in section 4. All the relevant RGEs for the Yukawa couplings, and gauge coupling constants are given in the Appendix A and RGEs for the neutrino oscillation parameters in Appendix B.
\section{Renormalization group equations}
  Renormalisation Group Equations (RGEs) \cite{da,rg1, beta} are studied in two different approaches as follows:

$\bullet$  top-down approach (a full theory is known at the high-energy scale and the theoretical predictions for neutrino parameters are given as initial conditions), and
 
$\bullet$ Bottom-up approach (starts with the experimental data of neutrino parameters at a low energy scale, and evolve them by using the RGEs at the high energy scale where a full theory of neutrino masses and lepton flavor mixing may exist).

\subsection{Bottom-up approach}

 In bottom-up approach, we evaluate the values of gauge and Yukawa couplings at high energy scale using RGEs. We can be divided into three regions, $ m_Z < \mu < m_t$, $m_t < \mu <m_s$, and $m_s< \mu <M_R$, using recent experimental data \cite{c2,pgd} as initial input values at low energy scale which are given in Table \ref{tab:1z}. 
 
\begin{table}[H]
\begin{adjustbox}{width=0.99\textwidth}
\centering
\begin{tabular}{lll}
\hline
Mass(GeV)& Coupling constants & Weinberg mixing angle \\ 
\hline
$m_Z(m_Z)$= 91.1876& $\alpha_{em}^{-1}(m_Z)$ =127.952 $\pm$ 0.009 & $\sin^2\theta_W(m_Z)$ = 0.23121 $\pm 0.00017$\\
 $m_t(m_t)$= 172.76 & $\alpha_{s}(m_Z)$ =0.1179$\pm$ 0.009\\
 $m_b(m_b)$= 4.18 &&\\
 $m_{\tau}(m_{\tau})$=1.77&&\\
\hline 
\end{tabular} 
\end{adjustbox}
\caption{ \footnotesize{Current experimental data for fermion masses, gauge coupling constants and Weinberg mixing angle.}}
\label{tab:1z}
\end{table}

From the relation, $\displaystyle \sin^2 \theta_{W}(m_{Z}) = \alpha_{em}(m_Z)/\alpha_2(m_Z)$, and using matching condition, 
\begin{equation}
\frac{1}{\alpha_{em}(m_Z)} =\frac{5}{3}\frac{1}{\alpha_1(m_Z)} + \frac{1}{\alpha_2(m_Z)},
\label{b}
\end{equation}
we calculated the values of gauge couplings, $\alpha_{2}$  for $SU(2)_{L}$ and $\alpha_{1}$ for $U(1)_{Y}$,

The normalised couplings \cite{da}, $g_i = \sqrt{4\pi \alpha_i}$, where  $\alpha_i$'s are the gauge couplings and $i=1, 2, 3$ denote electromagnetic, weak and strong couplings respectively. For the evolution from $m_Z$ scale to $m_t$ scale, we use the one-loop gauge couplings RGEs  \cite{bb1},
\begin{equation}
\frac{1}{\alpha_i(\mu)} = \frac{1}{\alpha_{i}(m_Z)} - \frac{b
_i}{2\pi}ln\frac{\mu}{m_Z},
\label{d}
\end{equation}
where $m_Z\leq \mu \leq m_t$  and $b_i$ = (5.30, -0.50, -4.00) for non-SUSY case. By using QED-QCD rescaling factor $\eta$ \cite{qcd}, fermion masses at $m_t$ scale is given by $\displaystyle m_b(m_t) = m_b(m_b)/\eta_b$ and $\displaystyle  m_{\tau}(m_t) = m_{\tau}(m_\tau)/\eta_{\tau}$, where $\displaystyle  \eta_{b} = 1.53$ and $\eta_{\tau} = 1.015$. The Yukawa couplings at $m_t$ scale  is given by 

\begin{equation}
\left.
\begin{array}{l}
\displaystyle  h_t(m_t) = m_t(m_t)/v_0 \\
\displaystyle  h_b(m_t) = m_b(m_b)/v_0 \eta_b \\
 \displaystyle  h_\tau(m_\tau) = m_\tau(m_\tau)/v_0\eta_\tau 
\end{array}
\right\},
\label{eq:t1}
\end{equation}
where the vacuum expectation value (vev), $v_0$ = 174 GeV is SM Higgs field. The calculated numerical values for fermion masses, Yukawa and gauge couplings at $m_t$ scale are given in Table \ref{tab:1b}.

\begin{table}[H]
\centering
\begin{tabular}{ccc} 
\hline
Fermions masses & Yukawa Couplings & Gauge Couplings \\
\hline
$m_t(m_t)$ = 172.76 GeV& $h_t(m_t)$ = 0.9928&$g_1(m_t)$ =  0.4635\\
$m_b(m_t)$ = 2.73 GeV & $h_b(m_t)$ = 0.0157&$g_2(m_t)$ = 0.6511\\
$m_{\tau}(m_t)$ = 1.75 GeV  & $h_{\tau}(m_t)$ = 0.0100&$g_3(m_t)$ = 1.1890\\
\hline
\end{tabular} 
\caption{\footnotesize{ Evaluated values for fermion masses, Yukawa  and gauge couplings at $m_t$ scale.}}
\label{tab:1b}
\end{table}
 
The evolution of gauge and Yukawa couplings for running from $m_t$ to the $M_R$ scale using RGEs which are given in Appendix A. The following matching conditions are applied at the transition point from SM ($m_t<\mu<m_s$) to MSSM ($m_s<\mu<M_R$)  at $m_s$ scale, as

\begin{equation}
\left.
\begin{array}{l}
g_i(SUSY) = g_i(SM) \\
    h_t(SUSY) = \frac{h_t(SM)}{\sin\beta}= h_t(SM) \times \frac{\sqrt{1+ \tan^2\beta}}{\tan\beta}\\
h_b(SUSY) = \frac{h_b(SM)}{\cos\beta}= h_b(SM) \times \sqrt{1+ \tan^2\beta}\\
h_\tau(SUSY) = \frac{h_\tau(SM)}{\cos\beta}= h_\tau(SM) \times \sqrt{1+ \tan^2\beta} 
\end{array}
\right\}
\label{eq:t1}
\end{equation}
   
     In this paper, for different input values of  $\tan\beta$, the behavior of $h_t$ and $h_b$ changes with the increase of $m_s$ scale, because they depends on $\tan\beta$ as shown in Eq. (\ref{eq:t1}) which are shown in Tables \ref{tab:1c} - \ref{tab:2c}, which will be further needed as initial inputs at the next top-down approach at high energy scale $M_R$.
 We evaluated the values of gauge and Yukawa couplings at two different high energy scales using RGEs at different values of $\tan \beta$ as given in Tables \ref{tab:1c} and \ref{tab:2c}.

\begin{table}[H]
\centering
\begin{tabular}{ccccc|ccc}
\hline 
$m_s$(TeV) &&$h_t$ &$h_b$ &$h_\tau$ &$g_1$ &$g_2$& $g_3$\\
\hline
2& &0.5884& 0.1917&   0.2449 &    0.6591&    0.7024 &  0.7357 \\ 
4& & 0.5753&   0.1912&   0.2471 &0.6548&      0.6980&  0.7328 \\ 
6 & & 0.5662&    0.1908&   0.2488&     0.6514&    0.6945&  0.7305  \\ 
8& $\tan\beta = 30$ & 0.5621&    0.1907&    0.2496&   0.6498&    0.6927&  0.7293 \\ 
10 &&   0.5582&0.1906&    0.2505&     0.6481&   0.6910&  0.7282 \\  
12 & & 0.5545&   0.1904&   0.2513&   0.6464&   0.6893&  0.7271\\
14 & & 0.5528&   0.1903&  0.2517&    0.6456&   0.6885&  0.7265 \\
\hline 
2& & 0.5806&0.1206&  0.1545 &    0.6592&    0.7025 &  0.7358 \\ 
4& & 0.5684&   0.1207&   0.1562 &0.6549&      0.6981&  0.7329 \\ 
6 && 0.5600&    0.1207&   0.1575&     0.6515&    0.6946&  0.7305  \\ 
8&$\tan\beta = 20$  & 0.5562&    0.1208&    0.1581&   0.6498&    0.6929&  0.7294 \\ 
10 & &  0.5526&0.1208&    0.1588&     0.6482&   0.6912&  0.7283 \\  
12 &  &0.5492&   0.1209&   0.1594&   0.6465&   0.6895&  0.7271\\
14 && 0.5476&   0.1209&  0.1597&    0.6457&   0.6886&  0.7266 \\
\hline 
\end{tabular} 
\caption{ \footnotesize{Evaluated values of Yukawa and gauge couplings  at $M_R =10^{15}$ GeV for different values of $\tan\beta$ at different choices of $m_s$ scale.}}
\label{tab:1c}
\end{table}

\begin{table}[H]
\centering
\begin{tabular}{ccccc|ccc}
\hline 
$m_s$(TeV) &&$h_t$ &$h_b$ &$h_\tau$ &$g_1$ &$g_2$& $g_3$\\
\hline
2&& 0.6107&  0.2030&    0.2515&   0.6329&    0.6967 &  0.7533 \\ 
4&& 0.5974&    0.2024&    0.2536 & 0.6291&       0.6923&  0.7502 \\ 
6 &&0.5882&     0.2019&    0.2553&     0.6261&    0.6889&  0.7477  \\ 
8&$\tan\beta$ = 30 &  0.5840&   0.2018&    0.2561&   0.6246&   0.6873&  0.7465\\ 
10 &&  0.5801&0.2016&     0.2569&     0.6231&   0.6856&  0.7452 \\  
12 &&  0.5763&  0.2014&   0.2577&   0.6217&   0.6839&  0.7440\\
14 &&  0.5745&  0.2013&  0.2581&    0.6209&   0.6831&  0.7434 \\
\hline 
2& & 0.6032&  0.1281&    0.1591&   0.6330&    0.6968 &  0.7534 \\ 
4&& 0.5974&    0.1281&    0.1608& 0.6292&       0.6925&  0.7502 \\ 
6 &&0.5824&     0.1281&    0.1621&     0.6262&    0.6891&  0.7478  \\ 
8&$\tan\beta$ = 20 &  0.5785&   0.1281&    0.1627&   0.6247&   0.6874&  0.7465\\ 
10 &&  0.5748&0.1282&     0.1634&     0.6232&   0.6857&  0.7453 \\  
12 & & 0.5713&  0.1282&   0.1640&   0.6217&   0.6840&  0.7441\\
14 &  &0.5696&  0.1282&  0.1643&    0.6210&   0.6832&  0.7435 \\
\hline
\end{tabular} 
\caption{ \footnotesize{Evaluated values of Yukawa and gauge couplings at $M_R =10^{14}$ GeV for different values of $\tan\beta$  at different choices of $m_s$ scale.}}
\label{tab:2c}
\end{table}

\subsection{Top-down Approach}
In top-down approach, we use the values of Yukawa and guage couplings at high energy scale $M_R$, as initial inputs. We consider a sc relation among mixing angles, $\theta_{23} = q \times ( \theta_{12} + \theta_{13})$, taking q = 1.11. In the present work, we express the neutrino mass eigenstates ($m_2$, $m_3$) in terms of  $m_1$ for NH as $m_2 = 1.236 \times m_1$ and $m_3 = 4.395 \times m_1$, considering $m_1 = 0.012$ eV. For IH we express the neutrino mass eigenstates ($m_1$, $m_2$) in terms of $m_3$ as $m_1 = 5.686 \times m_3$ and $m_2 = 5.731 \times m_3$, taking $m_3 = 0.009$ eV, where the sum of three neutrino masses $|\Sigma m_i|$ and $|\Delta m^{2}_{ij}|$ are in favor with the latest cosmological bound \cite{g,h} and latest experimental data respectively.  We fix the two Majorana phases $\psi_1$ and $\psi_2$ at  0 and 180, respectively and the dirac CP phase at $180^0$. Using all the necessary mathematical frameworks, we analyze the radiative nature of neutrino parameters like neutrino masses, mixings, phases, using top-down approach  with the variations of $m_s$ scale at fixed value of $\tan\beta$ for both NH and IH. In the present work, we also check the stability of SC relation among mixing angles. The input set at high energy scale is given in Table \ref{tab:2} and the latest experimental data, which we refer, in Table \ref{tab:3}. 

\begin{table}[H]
\centering
\begin{adjustbox}{width=0.5\textwidth}
\begin{tabular}{lll}

\hline 
input & NH&IH\\
\hline
$m_1$(eV)&0.012& $5.686 \times m_3$\\ 
$m_2$(eV)& 1.236 $\times m_1$& $5.731 \times m_3$ \\ 
$m_3$(eV)& 4.395$\times m_1$ &0.009\\ 
$\theta_{12}/^0$& 35.29&35.29\\ 
$\theta_{13}/^0$ &8.60 &8.60\\  
$\psi_1/^0$&0&0\\
$\psi_2/^0$ &180&180\\
$\delta_{CP}/^0$ &180&180\\
\hline 
\end{tabular} 
\end{adjustbox}
\caption{ \footnotesize{Input set of neutrino parameters at high energy scale for both NH and IH. We consider relations among the three neutrino mass eigenstates, $m_2 = 1.236 \times m_1$ and $m_3 = 4.395 \times m_1$ for NH, and  $m_1 = 5.686 \times m_3$ and $m_2 = 5.731 \times m_3$ for IH. Self-complementarity relation among the three mixing angles, $\theta_{23} = q\times (\theta_{13} + \theta_{12})$, with q = 1.11 is also used.}}
\label{tab:2}
\end{table}

\begin{table}[H]
\centering
\begin{adjustbox}{width=1.05\textwidth}
\begin{tabular}{cccc}

\hline Parameters & Best-fit  & 2$\sigma$& 3$\sigma$\\
\hline
$\Delta m^2_{21} [10^{-5} eV^2]$ &$ 7.50^{+0.22}_{-0.20}$  & 7.12 - 7.93& 6.94 - 8.14  \\
$\Delta m^2_{31} [10^{-3} eV^2]$ (NO) & $ 2.55^{+0.02}_{-0.03}$   & 2.49 - 2.60 & 2.47 - 2.63\\
$\theta_{23}(NO)$ & 49.26 $ \pm$ 0.79 & 47.37 - 50.71 & 41.20 - 51.33\\
$\theta_{12}$ & $ 34.3 \pm 1.0$&  32.3 - 36.4&  31.4 - 37.4 \\
$\theta_{13}(NO)$ & $ 8.53^{+0.13}_{-0.12}$ & 8.27 - 8.79 &8.17 - 8.96 \\
$\delta_{CP} (NO)$ &$ 194^{+24}_{-22}$ &  152 - 255 & 128 - 359\\
$|\Sigma m_i|$ &  $<$ 0.12 eV ; $\ge$ 0.06 eV &&\\
\hline 
\end{tabular} 
\end{adjustbox}
\caption{ \footnotesize{Current best-fit values, 1$\sigma$ errors, and 2$\sigma$ and 3$\sigma$ intervals for the neutrino oscillation parameters from global data which are adopted from References \cite{de,c2,pgd}.}}
\label{tab:3}
\end{table}

\section{Numerical Analysis  and Results}

\begin{figure}[H]
\centering
\begin{subfigure}{0.54\textwidth}
    \includegraphics[scale=0.78]{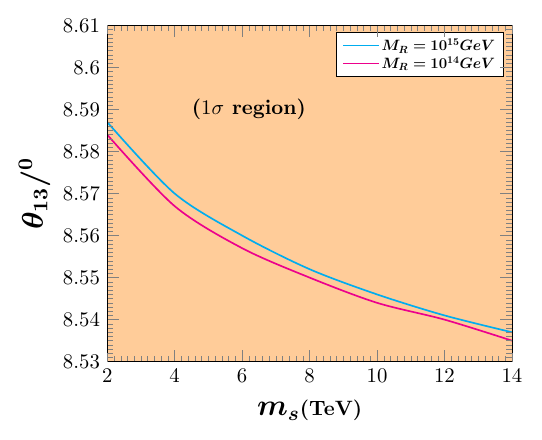}
\end{subfigure}
\begin{subfigure}{0.43\textwidth}
    \includegraphics[scale=0.78]{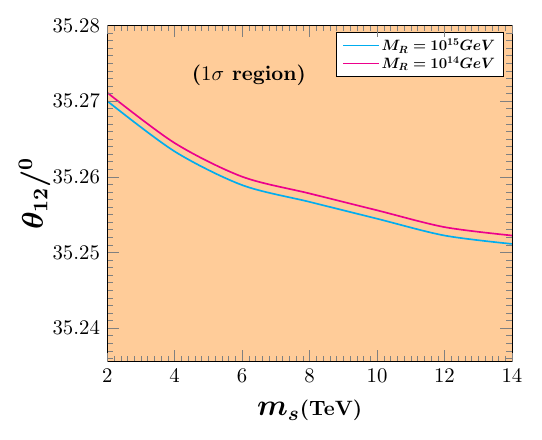}
\end{subfigure}
\begin{subfigure}{.54\textwidth}
    \includegraphics[scale=0.78]{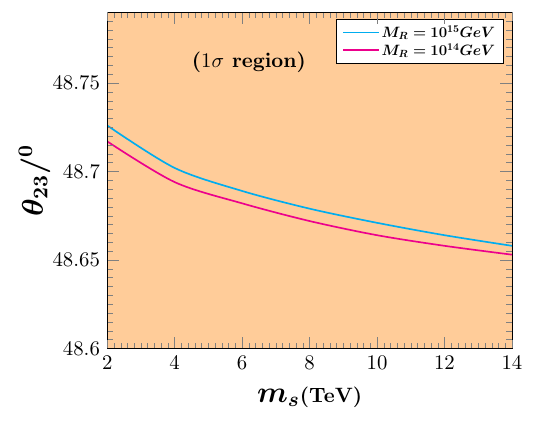}
\end{subfigure}
\begin{subfigure}{0.43\textwidth}
   \includegraphics[scale=0.78]{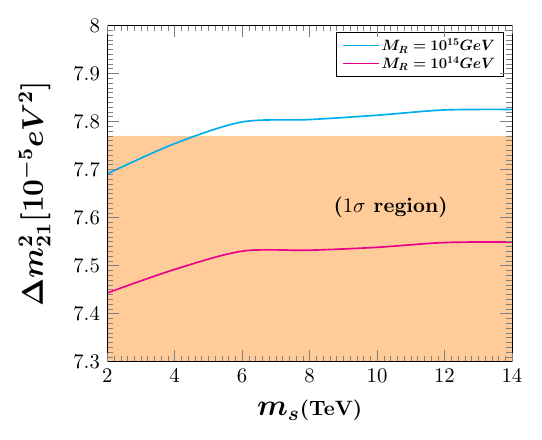}
\end{subfigure}
\begin{subfigure}{0.54\textwidth}
  \includegraphics[scale=0.78]{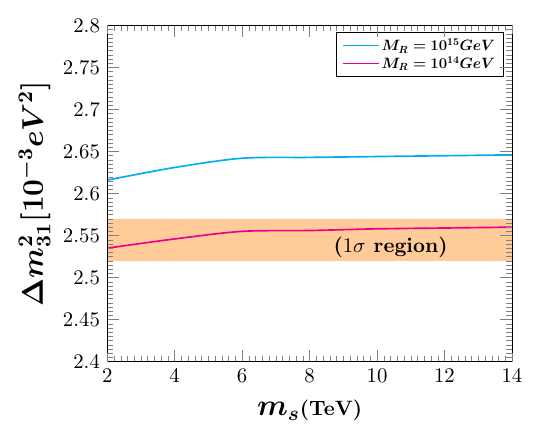}
\end{subfigure}
\begin{subfigure}{0.43\textwidth}
  \includegraphics[scale=0.78]{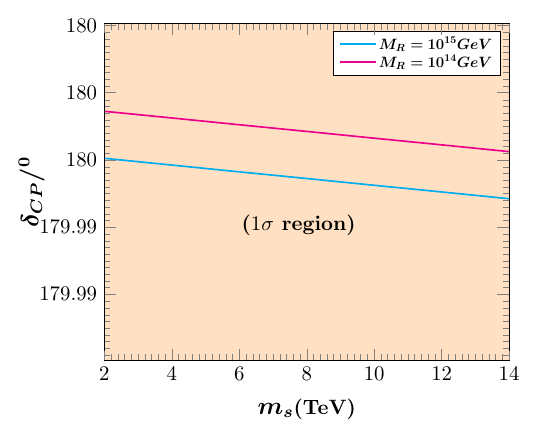}
\end{subfigure}

\caption{\footnotesize{For Normal Hierarchical (NH) case, effects on the low energy output results in $ \theta_{ij}$ and $ \Delta m^{2}_{ij}$  with the variations of $m_s$ at fixed value of $\tan \beta = 20$ with different values of $M_R$. The shaded portion is the $1\sigma$ region.}}
\label{fig:sfig1}
\end{figure}

\begin{figure}[H]
\centering
\begin{subfigure}{0.54\textwidth}
    \includegraphics[scale=0.78]{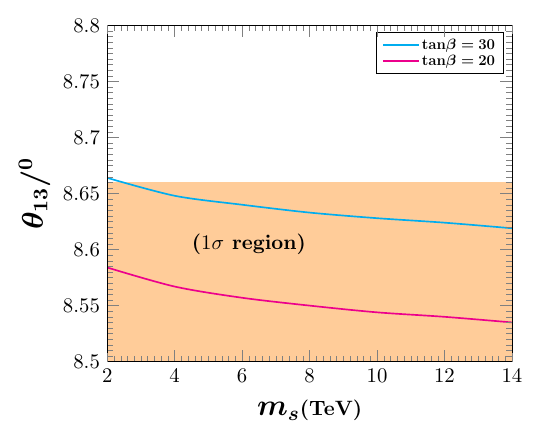}
\end{subfigure}
\begin{subfigure}{0.43\textwidth}
    \includegraphics[scale=0.78]{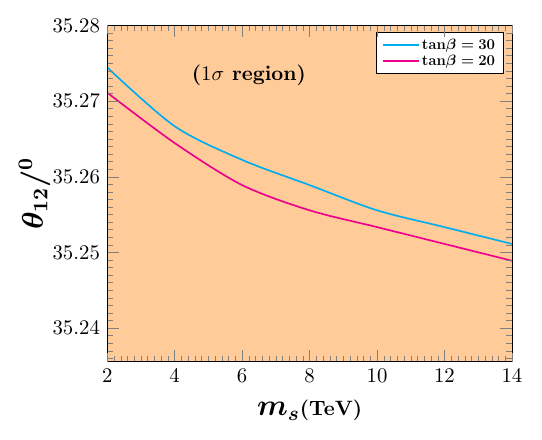}
\end{subfigure}
\begin{subfigure}{.54\textwidth}
    \includegraphics[scale=0.78]{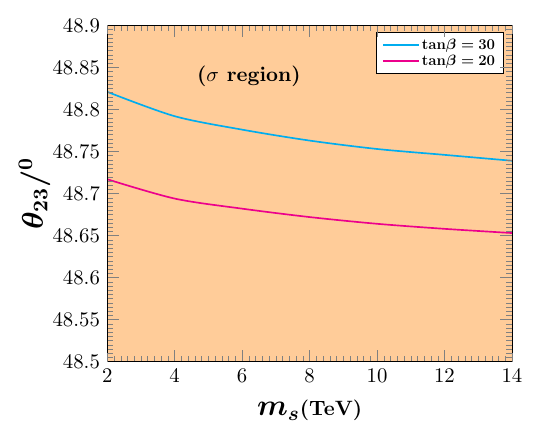}
\end{subfigure}
\begin{subfigure}{0.43\textwidth}
   \includegraphics[scale=0.78]{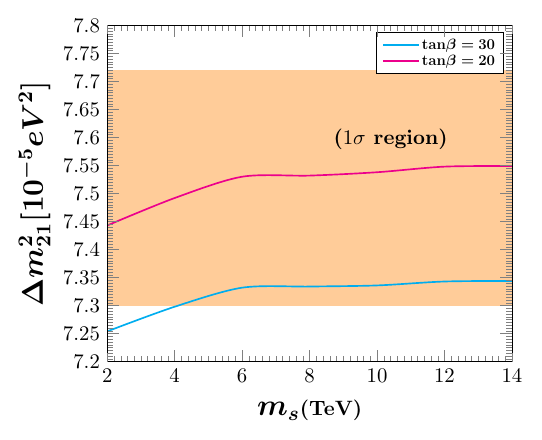}
\end{subfigure}
\begin{subfigure}{0.54\textwidth}
  \includegraphics[scale=0.78]{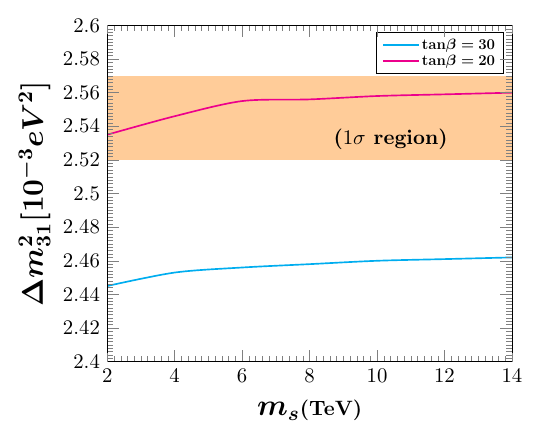}
\end{subfigure}
\begin{subfigure}{0.43\textwidth}
  \includegraphics[scale=0.78]{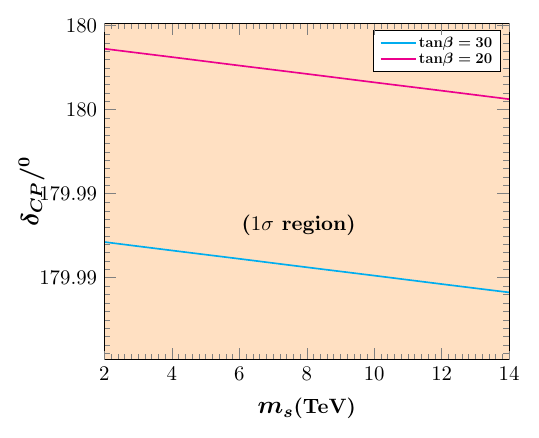}
\end{subfigure}

\caption{\footnotesize{For Normal Hierarchical (NH) case, effects on the low energy output results in $ \theta_{ij}$ and $ \Delta m^{2}_{ij}$  with the variations of $m_s$ at fixed value of $M_R = 10^{14}$ GeV with different values of $\tan \beta$. The shaded portion is the $1\sigma$ region.}}
\label{fig:sfig2}
\end{figure}

\begin{figure}[H]
\centering
\begin{subfigure}{0.54\textwidth}
    \includegraphics[scale=0.78]{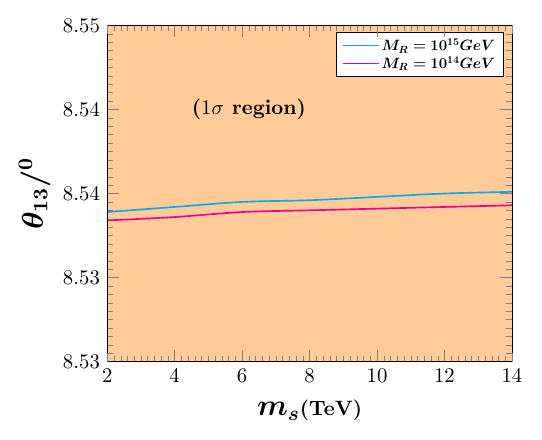}
\end{subfigure}
\begin{subfigure}{0.43\textwidth}
    \includegraphics[scale=0.78]{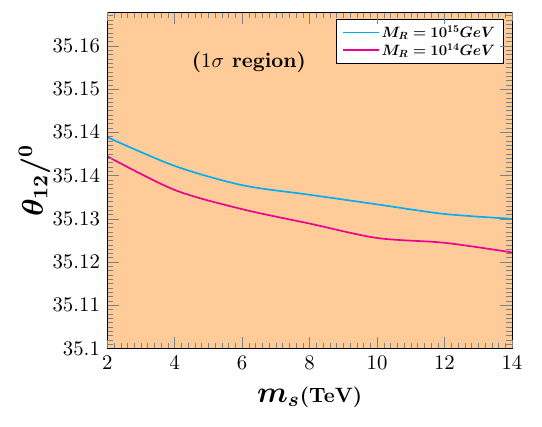}
\end{subfigure}
\begin{subfigure}{.54\textwidth}
    \includegraphics[scale=0.78]{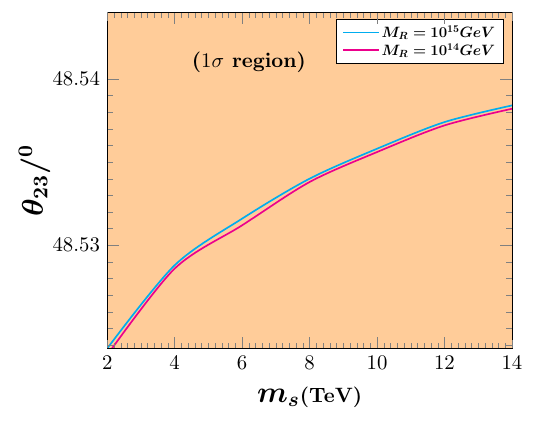}
\end{subfigure}
\begin{subfigure}{0.43\textwidth}
   \includegraphics[scale=0.78]{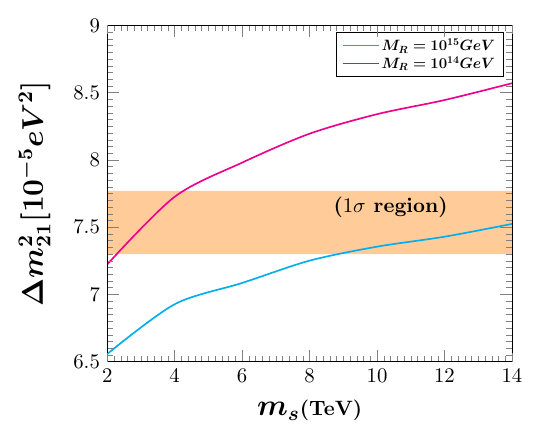}
\end{subfigure}
\begin{subfigure}{0.54\textwidth}
  \includegraphics[scale=0.78]{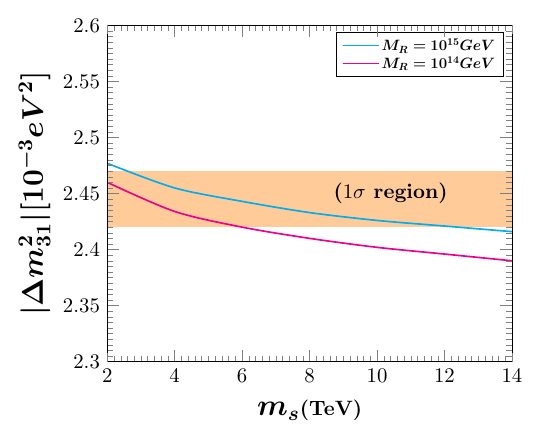}
\end{subfigure}
\begin{subfigure}{0.43\textwidth}
  \includegraphics[scale=0.78]{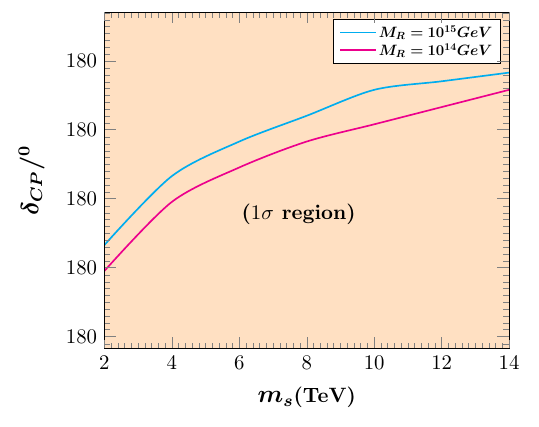}
\end{subfigure}

\caption{\footnotesize{For Inverted Hierarchical (IH) case, efffects on the low energy output results in $ \theta_{ij}$ and $ |\Delta m^{2}_{ij}|$  with the variations of $m_s$ at fixed value of $\tan \beta = 20$ with different values of $M_R$. The shaded portion is the $1\sigma$ region.}}
\label{fig:sfig3}
\end{figure}

\begin{figure}[H]
\centering
\begin{subfigure}{0.54\textwidth}
    \includegraphics[scale=0.78]{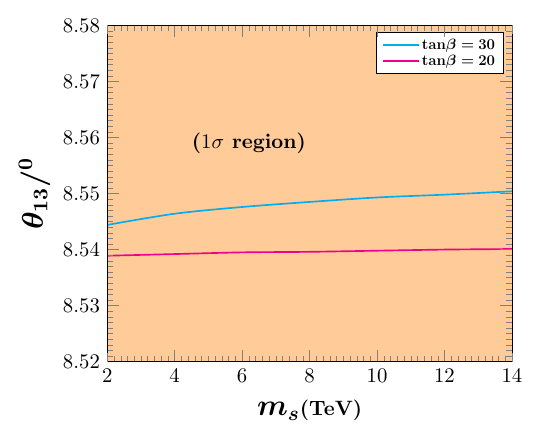}
\end{subfigure}
\begin{subfigure}{0.43\textwidth}
    \includegraphics[scale=0.78]{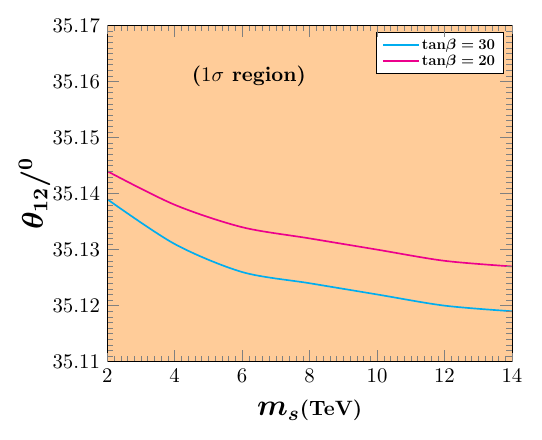}
\end{subfigure}
\begin{subfigure}{.54\textwidth}
    \includegraphics[scale=0.78]{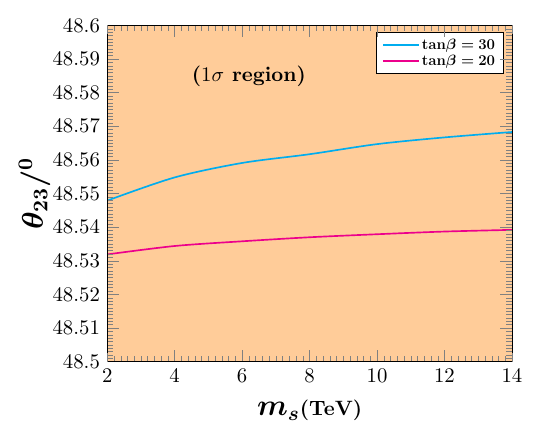}
\end{subfigure}
\begin{subfigure}{0.43\textwidth}
   \includegraphics[scale=0.78]{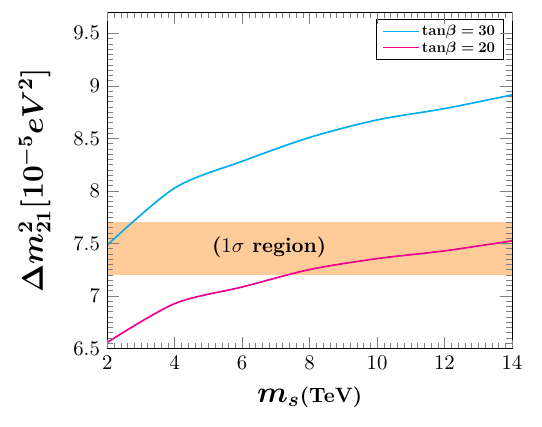}
\end{subfigure}
\begin{subfigure}{0.54\textwidth}
  \includegraphics[scale=0.78]{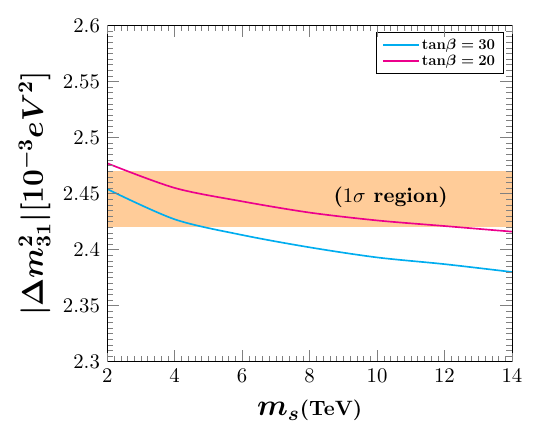}
\end{subfigure}
\begin{subfigure}{0.43\textwidth}
  \includegraphics[scale=0.78]{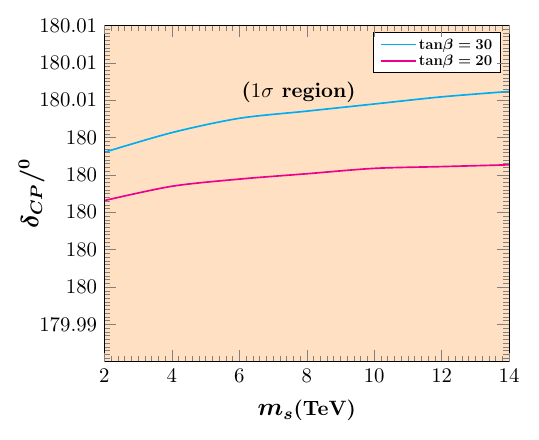}
\end{subfigure}

\caption{\footnotesize{For Inverted Hierarchical (IH) case, effects on the low energy output results in $ \theta_{ij}$ and $ |\Delta m^{2}_{ij}|$  with the variations of $m_s$ at fixed value of $M_R = 10^{15}$ GeV with different values of $\tan \beta$. The shaded portion is the $1\sigma$ region.}}
\label{fig:sfig4}
\end{figure}

In the present work, we take two different high energy scales, $M_R = 10^{15}$ GeV and $M_R = 10^{14}$ GeV at $\tan\beta$ = 30 and 20, respectively. We analyze  the effect of the  variation of $m_s$, $M_R$, and $\tan \beta$ for neutrino mass parameters and mixing angles. For both cases, the numerical data are given in Tables \ref{tab:4} -\ref{tab:5}, and its graphical representations in Figures \ref{fig:sfig1} -  \ref{fig:sfig4}. 
\textbf{•}
  For NH, high energy scale $M_R = 10^{14}$ GeV and $\tan \beta = 20$ are more preferred as the values of neutrino oscillation parameters are within $1\sigma$ range and all the neutrino oscillation parameters maintain stability with the variation of $m_s$ scale . For IH case, more higher energy scale $M_R = 10^{15}$ GeV and $\tan \beta = 20$  are preferred. From the numerical data, we observe that the mixing angles for NH increase more than the IH at the weak scale after renormalization group evolution. SC relation among mixing angles maintain stability at low energy scale with the variation of $m_s$ scale.

In the present work, we calculate the values of the effective light neutrino mass $m_{\beta \beta}$ and the neutrino mass extracted from ordinary beta decay $m_{\beta}$ at low energy scale. The absolute neutrino mass scale can be determined by the
following observations:

\begin{enumerate}[label=\roman*.] 
\item The formula for effective light neutrino mass $m_{\beta \beta}$ from neutrinoless double beta decay is given as follows: \cite{mbb1,mbb2} 

\begin{eqnarray}{\label{eqnarray:mbb}}
m_{\beta \beta} &=&  \biggl|\sum_{\mathclap{k=1}}^{3} |U_{ek}^{11}|^2 m_{k} \biggl|  \nonumber\\
 &=& \biggl|\cos ^2\theta_{12} \cos^2 \theta_{13} e^{ 2\iota \psi_1 }m_1 + \cos^2 \theta_{13} \sin^2 \theta_{12} e^{ 2\iota \psi_2} m_2 
+ \sin^2 \theta_{13} m_3\biggl|.
\end{eqnarray}

\item The formula for neutrino mass extracted from ordinary beta decay $m_{ \beta}$ is given as follows: \cite{mb1,mb2} 

\begin{eqnarray}{\label{eqnarray:mb}}
m_{\beta} &=& \sqrt{\sum_{\mathclap{k=1}}^{3} |U_{ek}^{11}|^2 m_{k}^2} \nonumber\\
 &=& \sqrt{\cos ^2\theta_{12} \cos^2 \theta_{13} m_1^2 + \cos^2 \theta_{13} \sin^2 \theta_{12} m_2^2 + \sin^2 \theta_{13} m_3^2},
\end{eqnarray}

\end{enumerate}
   where $|U_{ek}^{11}|^2$ are the elements of the PMNS matrix for mass value $m_k$ for k=1,2, and 3.
 The calculated values of effective Majorana  mass $m_{\beta \beta}$ at low energy scale  are found to lie within the following upper limits on the effective mass:   $m_{\beta \beta} <$ (104 - 228) meV by
Gerda \cite{gerda}, $m_{\beta \beta}  <$ (75 - 350) meV by CUORE \cite{cuore} and  $(m_{\beta \beta}^{min})_{IO} =  18.4 \pm 1.3 $ meV\cite{dinv} using Particle Data Group \cite {pgd}. The estimated values of neutrino mass extracted from ordinary beta decay $m_{\beta}$ at low energy are also found to lie within the bound given by KATRIN experiment  $m_{\beta \beta}  <$ 0.8 eV with 90 \% C.L. \cite {kato}. In this present work, it is found that both $m_{\beta \beta} $ and  $m_{\beta} $ increase with increasing $m_s$ for NH case whereas $m_{\beta \beta} $ increases but $m_{ \beta}$  decreases with increasing $m_s$ .

\begin{table}[H]
\centering
\begin{adjustbox}{width=1.05\textwidth}
\begin{tabular}{c|cccc|cccc|cccc}
\hline{$m_s$}& \multicolumn{4}{c|}{$\theta_{23}/^0$} & \multicolumn{4}{c|}{$\theta_{12}/^0$} & \multicolumn{4}{c}{$\theta_{13}/^0$}\\
(TeV) &A1 &A2 &B1&B2&A1 &A2 &B1&B2& A1 &A2 &B1&B2\\
\hline
2  &48.726 & 48.717 & 48.845 &48.821  &  35.271 & 35.272&35.276  &35.275 & 8.587  & 8.584 & 8.677 &8.664\\
4  & 48.702 & 48.694 & 48.816 & 48.792 & 35.265& 35.266 &35.269 & 35.268 & 8.570  &8.567 & 8.661 &8.648\\
6  & 48.689  & 48.682 &  48.799 &48.776 &   35.261 & 35.262 &35.265  &35.264 & 8.560 & 8.557 &8.652&8.640\\
8  & 48.679  & 48.672& 48.786&  48.763 &  35.259 & 35.260 &35.261 & 35.261  &8.552 & 8.550 & 8.644 &8.633\\
10 &48.671 & 48.664 &48.777 &  48.753 & 35.257& 35.258 &35.259  & 35.258  &8.546 & 8.544 & 8.639 &8.628\\
12 & 48.664 & 48.658 &48.769& 48.746 &  35.255 & 35.256 &35.257  & 35.256  &8.224 & 8.541 & 8.540 &8.624\\
14 &48.658  &48.653 & 48.762 & 48.739 &  35.254 & 35.255 &35.255  & 35.254  &8.537 & 8.535 &  8.630 &8.619\\
\hline 
$m_s$ &  \multicolumn{4}{c|}{$\Delta m^2_{31}$ ($10^{-3} eV^2$)} & \multicolumn{4}{c|}{$\Delta m^2_{21}$ ($10^{-5} eV^2$)}& \multicolumn{4}{c}{$|\Sigma m_i|$}\\
(TeV) &A1 &A2 &B1&B2&A1 &A2 &B1&B2& A1 &A2 &B1&B2\\
\hline 
 2&  2.616 &  2.535  &  2.515 &2.445 & 7.691  & 7.443  &7.476 & 7.254& 0.0781 & 0.0768  &0.0765 & 0.0754\\
 4&  2.631 & 2.546 & 2.527 &2.453  &   7.754 &  7.492 &7.534 & 7.298&0.0780 & 0.0779 &0.0764 & 0.0752\\
 6 & 2.642 & 2.555   & 2.534 &2.456 &  7.799 &  7.530  &7.578 & 7.332 & 0.0779  & 0.0766  &0.0763 & 0.0751\\
 8  & 2.643 & 2.556   &2.536&2.458 &  7.804 &  7.532  &7.579  &7.334&0.0778 & 0.0765 &0.0762 &0.0750\\
 10 & 2.644 & 2.558  & 2.537 &2.460  &  7.813 &  7.538  &7.586  &7.336 & 0.0777& 0.0764 &0.0761  &0.0749\\
 12 &2.645  & 2.559 & 2.538 &2.461 & 7.824 &  7.548 &7.596  &7.343& 0.0776 & 0.0763  &0.0760 &0.0748\\
 14 & 2.646  & 2.560  & 2.539 &2.462 & 7.825  & 7.549  &7.598  &7.344&0.0775 &0.0762  &0.0758 &0.0747\\
\hline 
\end{tabular} 
\end{adjustbox}
\caption{ \footnotesize{For Normal Hierarchy (NH) case, effects on the output of $\theta_{ij}$ , $\Delta m^2_{ij}$ and $\Sigma m_i$ at low energy scale, on varying $m_s$ at different  values of $\tan\beta$ and $M_R$. The sets A1 and A2 represent the output values assigned to the neutrino oscillation parameters at $M_R = 10^{15}$ GeV and $M_R = 10^{14}$ GeV respectively at $\tan \beta = 20$. Similarly, B1 and B2 represent the output sets assigned to the neutrino oscillation parameters at $M_R = 10^{15}$ GeV and $M_R = 10^{14}$ GeV respectively at $\tan \beta = 30$.}}
\label{tab:4}
\end{table}

\begin{table}[H]
\centering
\begin{adjustbox}{width=1.05\textwidth}
\begin{tabular}{c|cccc|cccc|cccc}
\hline{$m_s$}& \multicolumn{4}{c|}{$\theta_{23}/^0$} & \multicolumn{4}{c|}{$\theta_{12}/^0$} & \multicolumn{4}{c}{$\theta_{13}/^0$}\\
(TeV) &A1 &A2 &B1&B2&A1 &A2 &B1&B2& A1 &A2 &B1&B2\\
\hline
2  & 48.531 &48.531& 48.547 &48.548  & 35.144 & 35.140 &35.139 & 35.134 & 8.538 & 8.538 & 8.544 & 8.543 \\
4  &  48.534 & 48.534 & 48.554 &48.555 & 35.138 & 35.133 & 35.131 & 35.126 &8.539 & 8.538 & 8.546 & 8.545\\
6  &  48.535 & 48.535 &  48.559& 48.560 &  35.134 &  35.129& 35.126  &35.121&8.539 & 8.538 & 8.547&8.546\\
8  & 48.537 & 48.536& 48.561&   48.563 &  35.132  & 35.126 &35.124 & 35.118 & 8.539 & 8.539 & 8.548 &8.547\\
10 &48.537  &48.537 &48.564 &   48.565 &  35.130 & 35.123 &35.122 & 35.115&8.539 & 8.539 & 8.549 &8.548\\
12 & 48.538 & 48.538 &48.566&  48.567 &  35.128 & 35.122 &35.120 & 35.113 &8.540 & 8.539 & 8.549 &8.548\\
14 &48.539 &48.539 & 48.568 & 48.569 &  35.127 & 35.120 & 35.119 & 35.111&8.540& 8.539 & 8.550 &8.549\\
\hline 
$m_s$ &  \multicolumn{4}{c|}{$\Delta m^2_{31}$ ($10^{-3} eV^2$)} & \multicolumn{4}{c|}{$\Delta m^2_{21}$ ($10^{-5} eV^2$)}& \multicolumn{4}{c}{$|\Sigma m_i|$}\\
(TeV) &A1 &A2 &B1&B2&A1 &A2 &B1&B2& A1 &A2 &B1&B2\\
\hline 
 2&  2.477 & 2.460  &  2.454 &2.435 & 6.557 & 7.223  &7.483 & 8.195& 0.1096& 0.1092  &0.1093 & 0.1089\\
 4&  2.455 &2.434 & 2.427 &2.405  &   6.927 & 7.726 &8.027 &  8.876&0.1091& 0.1086&0.1088 & 0.1083\\
 6 & 2.443 &2.420   & 2.413 &2.388 &  7.085 & 7.980  &8.281  & 9.231 & 0.1089 & 0.1084  &0.1086& 0.1080\\
 8  & 2.433 &2.410  &2.402&2.376 & 7.251 & 8.195  & 8.508  &9.516& 0.1087 &0.1081 &0.1084 &0.1078\\
 10 & 2.426 &2.402  & 2.393 &2.366  &  7.355 & 8.340 &8.675 &9.714& 0.1086& 0.1080 &0.1082 &0.1076\\
 12 &2.421& 2.396 & 2.387 &2.359 &7.429  & 8.445& 8.783   &9.862&  0.1085& 0.1078 &0.1078&0.1075\\
 14 &  2.416& 2.390 & 2.380 &2.352 &7.524 & 8.570 & 8.913  &10.02& 0.1084 &0.1077  &0.1077&0.1073\\
\hline 
\end{tabular} 
\end{adjustbox}
\caption{ \footnotesize{ For Inverted Hierarchical (IH) case, effects on the output of $\theta_{ij}$ , $|\Delta m^2_{ij}|$ and $|\Sigma m_i|$ at low energy scale, with the variation of $m_s$ at different values of $\tan\beta$ and $M_R$. The sets A1 and A2 represent the output values assigned to the neutrino oscillation parameters at  $M_R = 10^{15}$ GeV and $M_R = 10^{14}$ GeV respectively at $\tan \beta = 20$. Similarly, B1 and B2 represent the output sets assigned to the neutrino oscillation parameters at $M_R = 10^{15}$ GeV and $M_R = 10^{14}$ GeV respectively at $\tan \beta = 30$.}}
\label{tab:5}
\end{table}

\begin{table}[H]
\centering
\begin{adjustbox}{width=.95\textwidth}
\begin{tabular}{c|c|cccc|cccc}
\hline{$m_s$}&& \multicolumn{4}{c|}{$m_{\beta \beta} $ (eV)} & \multicolumn{4}{c}{$m_{\beta}$ (eV)}\\
(TeV) & &A1 &A2 &B1&B2&A1 &A2 &B1&B2\\
\hline
2  && 0.0311 &0.0307& 0.0263 &0.0265 & 0.0398 &0.0393 &0.0360 & 0.0360 \\
4  && 0.0320 & 0.0316& 0.0271 &0.0273& 0.0404 & 0.0398 &  0.0366& 0.0365 \\
6  & & 0.0325 & 0.0321 &   0.0276& 0.0277& 0.0408 & 0.0402 &  0.0370  & 0.0368\\
8  &NH& 0.0329 &0.0324&  0.0279& 0.0280 &  0.0410 & 0.0404 & 0.0372&  0.0370 \\
10 &&0.0331  &0.0327 &0.0282&  0.0282& 0.0412 & 0.0406 &0.0374 &   0.0372\\
12 && 0.0334 & 0.0329 &0.0284& 0.0284 &  0.0414 & 0.0407&0.0375 &  0.0373\\
14 && 0.0336 &0.0330 &0.0285 & 0.0286 &  0.0415 &  0.0408 & 0.0376&  0.0374\\
\hline 
2  && 0.0185 &0.0185& 0.0187 & 0.0187  & 0.0324 & 0.0323 &0.0325 & 0.0324 \\
4  &&  0.0185 & 0.0185& 0.0187 & 0.0189 & 0.0323 &0.0321 &0.0324& 0.0322 \\
6  &&  0.0186 & 0.0186 & 0.0189& 0.0190&  0.0322& 0.0320 & 0.0324  &0.0322\\
8  &IH& 0.0186  & 0.0186& 0.0190&  0.0190&  0.0322  & 0.0320  &0.0323& 0.0321 \\
10 && 0.0186  &0.0186&0.0191 &  0.0191 & 0.0321 & 0.0319 &0.0323 &0.0321\\
12 && 0.0186& 0.0186 &0.0191&   0.0191& 0.0321 & 0.0319&0.0323 & 0.0321 \\
14 && 0.0186 &0.0186 & 0.0191 & 0.0191&  0.0321 &0.0318 &0.0323 &0.0320\\
\hline 
\end{tabular} 
\end{adjustbox}
\caption{ \footnotesize{Effects on the output of the effective light neutrino mass $m_{\beta \beta}$ and the neutrino mass extracted from ordinary beta decay $m_{\beta}$ at low energy scale, on varying $m_s$ for NH and IH cases at different $\tan\beta$ and $M_R$. The sets A1 and A2 represent the output values assign to the neutrino oscillation parameters at  $M_R = 10^{15}$ GeV and $M_R = 10^{14}$ GeV, respectively at $\tan \beta = 20$. Similarly, B1 and B2 represent the output sets assign to the neutrino oscillation parameters at $M_R = 10^{15}$ GeV and $M_R = 10^{14}$ GeV, respectively at $\tan \beta = 30$}}
\label{tab:6}
\end{table}
 
\begin{figure}[H]
\centering
\begin{subfigure}{.54\textwidth}
    \includegraphics[scale=0.78]{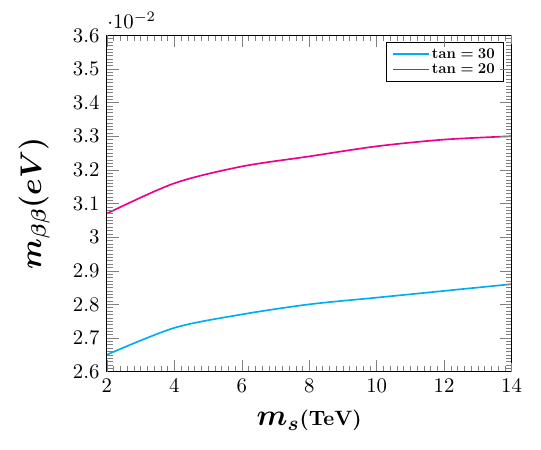}
\end{subfigure}
\begin{subfigure}{0.43\textwidth}
   \includegraphics[scale=0.78]{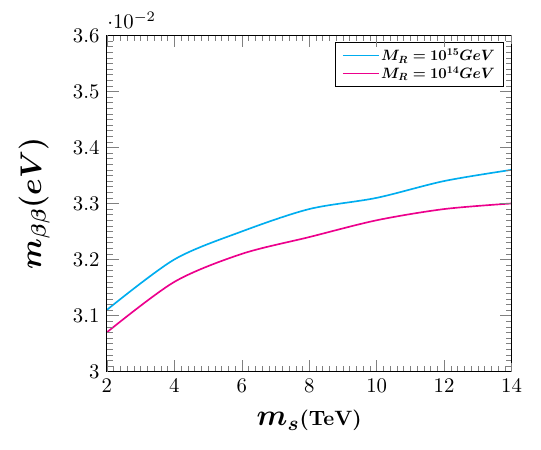}
\end{subfigure}
\begin{subfigure}{0.54\textwidth}
  \includegraphics[scale=0.78]{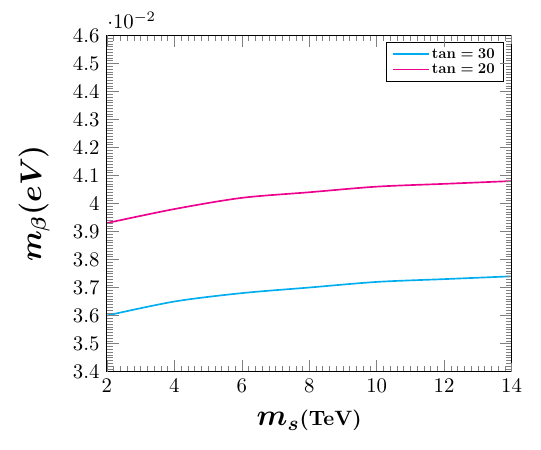}
\end{subfigure}
\begin{subfigure}{0.43\textwidth}
  \includegraphics[scale=0.78]{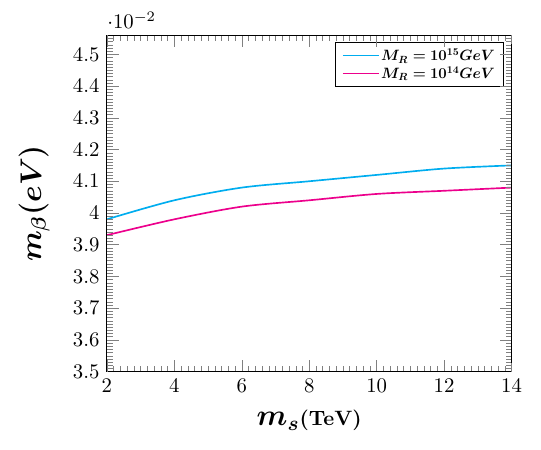}
\end{subfigure}

\caption{\footnotesize{For Normal Hierarchical (NH) case, effects on the low energy output results for the effective light neutrino mass $m_{\beta \beta}$ and the neutrino mass extracted from ordinary beta decay $m_{\beta}$ with the variations of $m_s$ at different values of $M_R$ and  $\tan \beta$.}}
\label{fig:sfig5}
\end{figure}


\begin{figure}[H]
\centering
\begin{subfigure}{.54\textwidth}
    \includegraphics[scale=0.78]{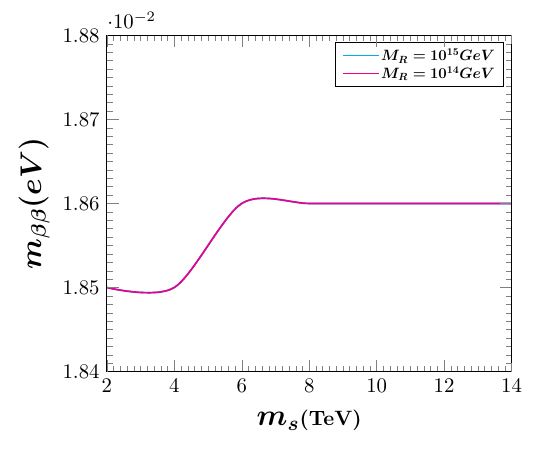}
\end{subfigure}
\begin{subfigure}{0.43\textwidth}
   \includegraphics[scale=0.78]{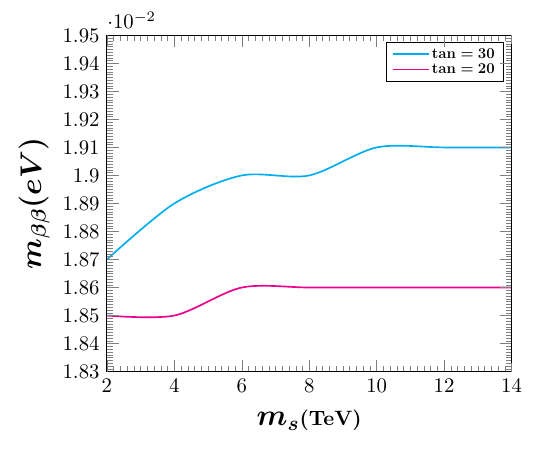}
\end{subfigure}
\begin{subfigure}{0.54\textwidth}
  \includegraphics[scale=0.78]{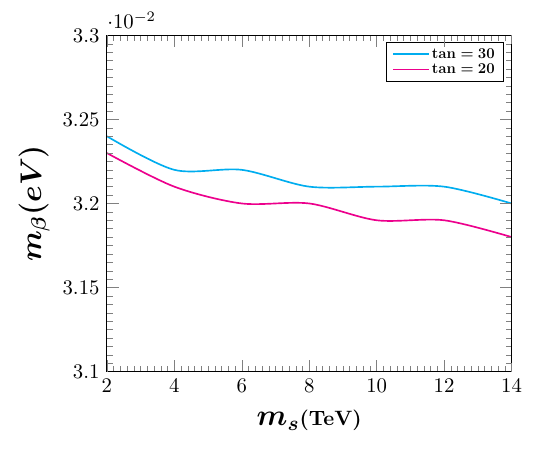}
\end{subfigure}
\begin{subfigure}{0.43\textwidth}
  \includegraphics[scale=0.78]{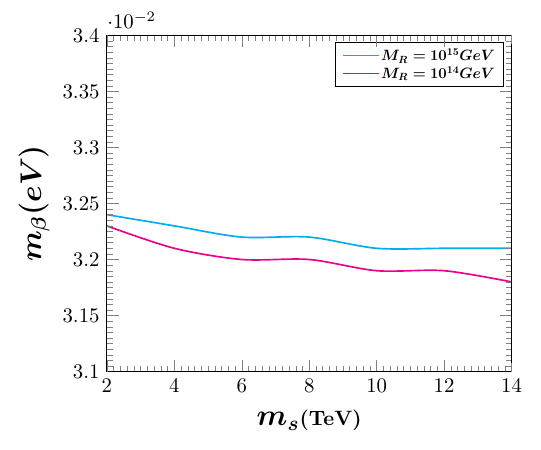}
\end{subfigure}

\caption{\footnotesize{For Inverted Hierarchical (IH) case, effects on the low energy output results for the effective light neutrino mass $m_{\beta \beta}$ and the neutrino mass extracted from ordinary beta decay $m_{\beta}$ with the variations of $m_s$ at different values of $M_R$ and  $\tan \beta$.}}
\label{fig:sfig5}
\end{figure}

\section{Discussion and Conclusions}
      The results are summarized as below.
 \begin{enumerate}[label=\roman*.] 
 \item All the mixing angles $\theta_{ij}$ are found to be decreased with increasing $m_s$ (2 TeV - 14 TeV) for NH, but increased for IH (except $\theta_{12}$) at different values of $\tan\beta$, and $M_R$ with the variation of $m_s$. $\Delta m_{ij}^2$  increases with increasing $m_s$ for NH, whereas $\Delta m_{31}^2$ decreases but $\Delta m_{21}^2$ increases for IH. All the neutrino oscillation parameters maintain stability with the variation of $m_s$ under RGEs for both cases.
 
 \item Our numerical outputs of neutrino oscillation parameters are consistent with the recent data within $1\sigma$ range at smaller $\tan \beta$ and higher $m_s$ for both cases. IH prefers higher energy scale $M_R = 10^{15}$ GeV,  whereas NH prefers smaller energy scale $M_R = 10^{14}$ GeV. NH is found to be more stable than IH.
 
 \item The SC relation among mixing angles is found to be consistent within $1\sigma$ range for all values of $\tan\beta$  and $M_R$.
 
 \item An interesting feature in this work is that mixing angles  at high energy scale for NH are found to be increased more than IH case at the weak scale after RG evolution.
 
 \item Our output values of $m_{\beta \beta}$ and $m_{\beta}$ at low energy scale for NH and IH cases with the variation of $m_s$ are found to be consistent with the latest experimental data for different values of $\tan \beta$ and $M_R$.
 \end{enumerate}

In this present work, the neutrino oscillation parameters for NH case at low energy scale including SC relations among mixing angles under radiative corrections with the variation of $m_s$ are found to be more stability than IH case. All the numerical values related to the absolute neutrino masses viz., $\Sigma |m_i|$, $m_{\beta}$ and $m_{ \beta \beta}$ are found to lie below the observational upper bound.
\section*{APPENDIX A}
\subsubsection*{RGEs for gauge couplings \cite{rg1}:}
The two loop RGEs for gauge couplings  are given by
\begin{equation}
 \frac{dg_i}{dt}= \frac{b_i}{16 \pi^2} g_i^3  + \frac{1}{(16 \pi^2)^2}\biggl[ \sum_{\mathclap{j=1}}^{3}b_{ij} g_i ^3 g_j ^2 - \sum_{\mathclap{j=t, b, \tau}} a_{ij}g_i ^3 h_j ^2 \biggl],
\end{equation} 

\bigskip
where t = ln $\mu$ and $ b_i ,b_{ij}, a_{ij}$ are $\beta$ function coefficients in MSSM, 
$$ b_{i} =
\left(\begin{array}{ccc}
6.6,  & 1.0,  &   -3.0
\end{array}\right) ,
b_{ij}=
\left(\begin{array}{ccc}
7.96 & 5.40 & 17.60 \\ 
1.80 & 25.00 & 24.00 \\ 
2.20 & 9.00 & 14.00
\end{array}\right),$$
$$ a_{ij}=
\left(\begin{array}{ccc}
5.2 & 2.8 & 3.6 \\ 
6.0 & 6.0 & 2.0 \\ 
4.0 & 4.0 & 0.0
\end{array}\right)$$ 
 and, for non-supersymmetric case, we have\\
 
$$ 
b_i=
\left(\begin{array}{ccc}
4.100,  & -3.167,  &   -7.00
\end{array}\right) ,
g_{ij}=
\left(\begin{array}{ccc}
3.98 & 2.70 & 8.8 \\ 
0.90 & 5.83 & 12.0 \\ 
1.10 & 4.50 & -26.0
\end{array}\right),$$  and 
$$a_{ij}=
\left(\begin{array}{ccc}
0.85 & 0.5 & 0.5 \\
1.50 & 1.5 & 0.5 \\ 
2.00 & 2.0 & 0.0
\end{array}\right).$$  \\

\subsubsection*{Two-loop RGEs for Yukawa couplings and quartic Higgs coupling \cite{rg1}:} 
  For MSSM,
\begin{eqnarray}
\frac{dh_t}{dt} &=&\frac{h_t}{16 \pi^2}\biggl(6h_{t}^2 + h_{b}^2 -\sum_{\mathclap{i=1}}^{3} c_i g_{i}^2 \biggl)
+\frac{h_t}{(16 \pi^2)^2}\biggl[\sum_{\mathclap{i=1}}\biggl(c_i b_i +\frac{c_{i}^2}{2}\biggl)g_{i}^4  + g_{1}^2 g_{2}^2 \biggl. \nonumber\\
& & + \frac{136}{45} g_{1}^2 g_{3}^2 + 8 g_{2}^2 g_{3}^2 +\biggl(\frac{6}{5}g_{1}^2+6 g_{2}^2+16g_{3}^2\biggl)h_{t}^2+ \frac{2}{5}g_{1}^2 h_{b}^2 -22h_{t}^4 \nonumber\\
&& \biggl.- 5 h_{b}^4-5h_{t}^2h_{b}^2 - h_{b}^2 h_{\tau}^2\biggl],
\end{eqnarray}

\begin{eqnarray}
\frac{dh_b}{dt} &=&\frac{h_b}{16 \pi^2}\biggl(6h_{b}^2 + h_{\tau}^{2}+ h_{t}^2- \sum_{\mathclap{i=1}}^{3} c_{i}^{'} g_{i}^2\biggl)+\frac{h_{b}}{(16 \pi^{2})^2}\biggl[ \sum_{\mathclap{i=1}} \biggl(c_{i}^{'} b_i +\frac{c{i}^{'2}}{2}\biggl) g_{i}^4 \biggl. \nonumber\\
&&+ g_{1}^2 g_{2}^2 +\frac{8}{9}g_{1}^2 g_{3}^2  + 8g_{2}^2 g_{3}^2+\biggl(\frac{2}{5} g_{1}^2 +6g_{2}^2+16 g_{3}^2\biggl) h_{b}^2 +\frac{4}{5} g_{1}^2 h_{t}^2 + \frac{6}{5}g_{1}^2 h_{\tau}^2   \nonumber\\
&& \biggl.-22h_{b}^4 - 3 h_{\tau}^4-5h_{t}^4 -5h_{b}^2 h_{t}^2 -3h_{b}^2 h_{\tau}^2\biggl],
\end{eqnarray}

\begin{eqnarray}
\frac{dh_\tau}{dt} &=&\frac{h_\tau}{16 \pi^2}\biggl(4h_{\tau}^2 +3 h_{b}^{2}- \sum_{\mathclap{i=1}}^{3} c_{i}^{''} g_{i}^2\biggl)+\frac{h_{\tau}}{(16 \pi^{2})^2}\biggl[ \sum_{\mathclap{i=1}} \biggl(c_{i}^{''} b_i +\frac{c{i}^{''2}}{2}\biggl)
 g_{i}^4 \biggl. \nonumber\\
&& \biggl.+\frac{9}{5} g_{1}^2 g_{2}^2 + \biggl(\frac{6}{5}g_{1}^2+6 g_{2}^2\biggl) h_{\tau}^2 +\biggl(\frac{-2}{5} g_{1}^2 +16g_{3}^2\biggl)h_{b}^2 +9 g_{b}^4 \biggl. \nonumber\\
&& \biggl.- 10 h_{\tau}^4-3 h_{b}^2 h_{t}^2  -9h_{b}^2 h_{\tau}^2\biggl],
\end{eqnarray}

where  $$c_{i}=
\left(\begin{array}{ccc}
\frac{13}{15}, & 3, & \frac{16}{13}
\end{array}\right) , 
c_{i}^{'}=
\left(\begin{array}{ccc}
\frac{7}{15}, & 3, & \frac{16}{3}  
\end{array}\right)$$  and
$$c_{i}^{''}=
\left(\begin{array}{ccc}
\frac{9}{5} ,& 3 ,& 0  
\end{array}\right).$$  
For non-supersymmetric case,

\begin{eqnarray}{\label{eq:t}}
\frac{dh_t}{dt} &=&\frac{h_t}{16 \pi^2}\biggl(\frac{3}{2} h_{t}^2 - \frac{3}{2} h_{b}^2 +Y_{2}(S) - \sum_{\mathclap{i=1}}^{3} c_i g_{i}^2\biggl) + \frac{h_t}{(16 \pi^2)^2}\biggl[\biggl(\frac{1187}{600}\biggl)g_{1}^4 - \frac{23}{4} g_{2}^4  \biggl. \nonumber\\
&&\biggl.- 108 g_{3}^4-\frac{9}{20} g_{1}^2 g_{2}^2+ \frac{19}{15} g_{1}^{2} g_{3}^2 +9 g_{3}^2 g_{2}^2 +\biggl(\frac{223}{80}g_{1}^2+\frac{135}{16} g_{2}^2 +16g_{3}^2\biggl)h_{t}^2  \biggl. \nonumber\\
&&-\biggl( \frac{43}{80}g_{1}^2- \frac{9}{16} g_{2}^2+ 16 g_{3}^2 \biggl) h_{b}^2+\frac{5}{2} Y_{4}(S) - 2 \lambda \biggl(3h_{t}^2+ h_{b}^2\biggl)+ \frac{3}{2} h_{t}^4-\frac{5}{4} h_{t}^2 h_{b}^2   \biggl. \nonumber\\
&&+\frac{11}{4} h_{b}^4+ Y_{2}(S)\biggl(\frac{5}{4}h_{b}^2 - \frac{9}{4}h_{t}^2\biggl)
- \eta _{4}(S)+\frac{3}{2}\lambda^2 \biggl],
\end{eqnarray}

\begin{eqnarray}{\label{eq:b}}
\frac{dh_b}{dt} &=&\frac{h_b}{16 \pi^2}\biggl(\frac{3}{2} h_{b}^2 - \frac{3}{2} h_{t}^2 +Y_{2}(S) - \sum_{\mathclap{i=1}}^{3} c_{i}^{'} g_{i}^2 \biggl)+\frac{h_b}{(16 \pi^2)^2}\biggl[ -\frac{127}{600}g_{1}^4-  \frac{23}{4} g_{2}^4 - 108 g_{3}^4 \biggl. \nonumber\\
&& \biggl. - \frac{27}{20} g_{1}^2 g_{2}^2+ \frac{31}{15} g_{1}^{2} g_{3}^2 +9 g_{3}^2 g_{2}^2 -\biggl (\frac{79}{80}g_{1}^2-\frac{9}{16} g_{2}^2 +16g_{3}^2\biggl)h_{t}^2 + \biggl( \frac{187}{80}g_{1}^2+ \frac{135}{16} g_{2}^2  \biggl. \nonumber\\
&&+ 16 g_{3}^2 \biggl) h_{b}^2 +\frac{5}{2} Y_{4}(S)- 2 \lambda \biggl(3h_{t}^2+3 h_{b}^2\biggl) + \frac{3}{2} h_{b}^4-\frac{5}{4} h_{t}^2 h_{b}^2 +\frac{11}{4} h_{t}^4  \biggl. \nonumber\\
 &&\biggl.+ Y_{2}(S)\biggl(\frac{5}{4}h_{t}^2 - \frac{9}{4}h_{b}^2\biggl) - \eta _{4}(S)+\frac{3}{2}\lambda^2 \biggl],
\end{eqnarray}

\begin{eqnarray}{\label{eq:tau}}
\frac{dh_\tau}{dt} &=&\frac{h_\tau}{16 \pi^2}\biggl(\frac{3}{2} h_{\tau}^2 + Y_{2}(S) - \sum_{\mathclap{i=1}}^{3} c_i^{''}g_{i}^2 \biggl)
+\frac{h_\tau}{(16 \pi^2)^2}\biggl[ \frac{1371}{200}g_{1}^4-  \frac{23}{4} g_{2}^4 -  \frac{27}{20} g_{1}^2 g_{2}^2 \biggl. \nonumber\\
&& \biggl.+\biggl(\frac{387}{80}g_{1}^2+ \frac{135}{16} g_{2}^2 \biggl)h_{\tau}^2 +\frac{5}{2} Y_{4}(S) - 6\lambda h_{t}^2+ \frac{3}{2}h_{\tau}^4 - \frac{9}{4}Y_{2}(S)h_{\tau}^2 \biggl. \nonumber\\
&&\biggl.- \eta _{4}(S) +\frac{3}{2} \lambda^2\biggl],
\end{eqnarray}

\begin{eqnarray}
\frac{d\lambda}{dt} &=&\frac{1}{16 \pi^2}\biggl[\frac{9}{4}\biggl(\frac{3}{25} g_{1}^4 + \frac{2}{5} g_{2}^2 g_{1}^2 + g_{2}^4 \biggl)-\biggl(\frac{9}{5} g_{1}^2 +9 g_{2}^2\biggl)\lambda 
+ 4 Y_{2}(S) \lambda -4H(S) +12 \lambda^2\biggl]
\biggl. \nonumber\\
&&\biggl.+\frac{1}{(16 \pi^2)^2}\biggl[ -78 \lambda^3 + 18\biggl(\frac{3}{5}g_{1}^2+ 3  g_{2}^2 \biggl)\lambda^2 +\biggl(- \frac{73}{8} g_{2}^4 + \frac{117}{20} g_{1}^{2} g_{2}^2 + \frac{1887}{200}g_{1}^4\biggl)\lambda\biggl. \nonumber\\
&&\biggl.+\frac{305}{8} g_{2}^6  - \frac{867}{120}g_{1}^2 g_{2}^4- \frac{1677}{200} g_{1}^4 g_{2}^2-\frac{3411}{1000} g_{1}^6 - 64 g_{3}^2\biggl(h_{t}^4+ h_{b}^4\biggl)- \frac{8}{5} g_{1}^2\biggl(2 h_{t}^4 - h_{b}^4 \biggl. \nonumber\\
&&\biggl.+3 h_{\tau}^4 \biggl)- \frac{3}{2}g_{2}^4 Y_{2}(S)+10 \lambda Y_{4}(S)+ \frac{3}{5}g_{1}^2(- \frac{57}{10}g_{1}^2 + 21 g_{2}^2)h_{t}^2+\biggl(\frac{3}{2}g_{1}^2 +9 g_{2}^2\biggl)h_{b}^2 \biggl. \nonumber\\
&&\biggl. +\biggl(-\frac{15}{2} g_{1}^2 +11 g_{2}^2\biggl)h_{\tau}^2 -24 \lambda^2 Y_{2}(S)-
 \lambda H(S)+6 \lambda h_{t}^2 h_{b}^2 +20 \biggl(3h_{t}^6 +3 h_{b}^6+h_{\tau}^6\biggl)\biggl. \nonumber\\
 &&\biggl. -12\biggl(h_{t}^4 h_{b}^2+h_{t}^2 h_{b}^4\biggl)\biggl],
\end{eqnarray}
where 
 $$ Y_{2}(S)=3 h_{t}^2+ 3 h_{b}^2+ h_{\tau}^2 ,$$
$$ Y_{4}(S)=\frac{1}{3}\biggl[3 \Sigma c_{i} g_{i}^2 h_{t}^2+3 \Sigma c_{i}^{'} g_{i}^2 h_{b}^2+ 3 \Sigma c_{i}^{''} g_{i}^2 h_{\tau}^2\biggl],$$

$$ H(S)=3 h_{t}^4+ 3 h_{b}^4+ h_{\tau}^4, $$

 $$\eta_{4}(S)=\frac{9}{4} \biggl[3 h_{t}^4 + 3 h_{b}^4 + h_{\tau}^4 -\frac{2}{3} h_{t}^2 h_{b}^2\biggl]$$

and $\displaystyle \lambda=  \frac{m_{h}^2}{v_0^2}$ is the Higgs self-coupling, $ m_h$ = 125.78 $\pm $ 0.26 GeV  is the Higgs mass \cite{higgs} and  $v_0$ = 174 GeV is the vacuum expectation value.
\\

The beta function coefficients for  non-SUSY case are given below:

$ \displaystyle c_{i}=
\left(\begin{array}{ccc}
0.85, & 2.25, & 8.00
\end{array}\right), $ 
$\displaystyle c_{i}^{'}=
\left(\begin{array}{ccc}
0.25 & 2.25, & 8.00  
\end{array}\right)$
and
$\displaystyle c_{i}^{''}=
\left(\begin{array}{ccc}
2.25 ,& 2.25 ,& 0.00  
\end{array}\right).$
\section*{APPENDIX B}

\subsubsection*{ RGEs for three neutrino mixing angles and phases \cite{rg2}: (neglecting higher order of $\theta_{13}$)}

\begin{equation}{\label{eq:t12}}
\dot{\theta}_{12}= -\frac{Cy_{\tau}^2}{32 \pi^2} \sin2\theta_{12} s_{23}^2 \frac{|m_1 e^{i\psi_1} +m_2 e^{i\psi_2}|^2}{\Delta{m_{21}^2}},
\end{equation}
$$
\dot{\theta}_{13}= \frac{Cy_{\tau}^2}{32 \pi^2}\sin2\theta_{12}\sin2\theta_{23}\frac{m_3}{\Delta{m_{31}^2}(1+\xi)}  
$$
\begin{equation}{\label{eq:t13}}
 \times \biggl[ m_1 \cos(\psi_{1}-\delta) -(1+\xi)m_2 \cos(\psi_2 - \delta)- \xi m_3 \cos\delta \biggl],
\end{equation}
\begin{equation}
\dot{\theta}_{23}= -\frac{C y_{\tau}^2}{32 \pi^2} \sin2\theta_{23}\frac{1}{\Delta m_{31}^2}\left[ c_{12}^2 |m_2 e^{i\psi_{2}} +m_3|^2 + s_{12}^2 \frac{|m_1 e^{i\psi_1} +m_3|^2}{1+\xi}\right],
\end{equation}

 where $\Delta m_{21}^2 = m_2^2 - m_1 ^2$ , $\Delta m_{31}^2 = m_3^2 - m_1 ^2$ and $\xi = \frac{\Delta m_{21}^2 }{\Delta m_{31}^2 }$. 
\subsubsection*{ RGEs for the three phases \cite{rg2}:}
 For  dirac  phase $\delta$:
\begin{equation}
 \dot{\delta} = \frac{C y_{\tau}^2}{32 \pi^2} \frac{\delta^{(-1)}}{\theta_{13}} + \frac{C y_{\tau}^2}{8 \pi^2} \delta^{(0)}, 
\end{equation}
where
\begin{eqnarray}
\delta^{(-1)}& = &\sin 2 \theta_{12} \sin 2 \theta_{23}\frac{m_3}{\Delta m_{31}^2(1+\xi) } \times \biggl[ m_1 \sin(\psi_1- \delta) 
\biggl. \nonumber\\
&&\biggl.-(1+\xi) m_2 \sin(\psi_2 - \delta) + \xi m_3 \sin\delta\biggl], 
\end{eqnarray}

\begin{eqnarray}
\delta^{(0)} &=& \frac{m_1 m_2 s_{23}^2 \sin(\psi_1 - \psi_2)}{\Delta m_{21}^2} \biggl. \nonumber\\
&&\biggl.  +  m_3 s_{12}^2\biggl[ \frac{m_1 \cos 2\theta_{23} \sin\psi_1}{\Delta m_{31}^2(1+\xi)}+ \frac{m_2 c_{23}^2 \sin(2\delta - \psi_2)}{\Delta m_{31}^2} \biggl]\biggl. \nonumber\\
&&\biggl. +  m_3 c_{12}^2\biggl[ \frac{m_1 c_{23}^2 \sin(2 \delta - \psi_1)}{\Delta m_{31}^2(1+\xi)}+ \frac{m_2 \cos(2\theta_{23})\sin\psi_2}{\Delta m_{31}^2} \biggl].
\end{eqnarray}

For Majorana phase $\psi_1$ \cite{rg2}:
\begin{eqnarray}
\dot{\psi_1}&=&\frac{C y_{\tau}^2}{8 \pi^2} \left[ m_3 \cos 2 \theta_{23} \frac{m_1 s_{12}^2 \sin \psi_1 + (1+ \xi) m_2 c_{12}^2 \sin \psi_2}{\Delta m_{31}^2(1+ \xi)}\right]\biggl. \nonumber\\
 &&\biggl. + \frac{C y_{\tau}^2}{8 \pi^2}\left[ \frac{m_1 m_2 c_{12}^2 s_{23}^2 sin(\psi_1 - \psi_2)}{
\Delta m_{21}^2}\right] 
\end{eqnarray}
 For Majorana phase $\psi_2$:
\begin{eqnarray}
 \dot{\psi_2}&=&\frac{C y_{\tau}^2}{8 \pi^2}\left[  m_3 \cos 2 \theta_{23} \frac{m_1 s_{12}^2 \sin \psi_1 +(1+\xi) m_2 c_{12}^2 \sin \psi_2  }{\Delta m_{31}^2 (1+\xi)}\right]\biggl. \nonumber\\
 &&\biggl.+\frac{C y_{\tau}^2}{8 \pi^2}\left[ \frac{m_1 m_2 s_{12}^2 s_{23}^2 \sin(\psi_1 - \psi_2
)}{\Delta m_{21}^2} \right] 
\end{eqnarray}
\subsubsection*{RGEs for neutrino mass eigenvalues \cite{rg2}:}
\begin{equation}
\dot{m_1} = \frac{1}{16 \pi^2}\left[  \alpha + C y_{\tau}^2 (2 s_{12}^2 s_{23}^2 +F_1)\right]m_1,
\end{equation}
\begin{equation}
\dot{m_2} = \frac{1}{16 \pi^2}\left[  \alpha + C y_{\tau}^2 (2 c_{12}^2 s_{23}^2 +F_2)\right]m_2,
\end{equation}
\begin{equation}
\dot{m_3} = \frac{1}{16 \pi^2}\left[  \alpha + 2C y_{\tau}^2 c_{13}^2 c_{23} \right]m_3,
\end{equation}
where
\begin{equation}
 F_1 = - s_{13}\sin 2 \theta_{12} \sin 2 \theta_{23} \cos \delta + 2 s_{13}^2 c_{12}^2 c_{23}^2,
 \end{equation}
 \begin{equation}
 F_2 =  s_{13}\sin 2 \theta_{12} \sin 2 \theta_{23} \cos \delta + 2 s_{13}^2 s_{12}^2 s_{23}^2.
 \end{equation}
  
For MSSM case:
  
$$\alpha= -\frac{6}{5} g_{1}^2 - 6 g_2 ^2 +6 y_t^2 $$ and $$C=1.$$

For SM case: 
 
$$\alpha= -3 g_{2}^2 + 2 y_{\tau}^2  +6 y_{t}^2 +6 y_{b}^2 + \lambda,$$
$$C=-\frac{3}{2}$$ and $\lambda$ is the Higgs self-coupling in the SM.

\section*{Acknowledgements}
One of the author (KHD) would like to thank Manipur University for financial support. 
\bibliographystyle{ieeetr}
\bibliography{refg}
\end{document}